\documentclass{article}
\usepackage[utf8]{inputenc}
\usepackage[T1]{fontenc}
\usepackage{siunitx}
\usepackage{xcolor}
\usepackage{soul}
\usepackage{caption}
\usepackage[english]{babel}
\usepackage{graphicx} 
\usepackage{lipsum} 
\usepackage[a4paper,top=3cm,bottom=2cm,left=3cm,right=3cm]{geometry} 
\usepackage[fontsize=13pt]{scrextend} 
\usepackage{sectsty} 
\usepackage{titlesec}
\usepackage{caption} 
\usepackage{amsmath} 
\usepackage{esint} 
\usepackage{tabularx} 
\usepackage{array} 
\usepackage{float} 
\usepackage{subcaption} 
\usepackage{hyperref} 
\usepackage{comment} 
\usepackage{geometry}
\usepackage{afterpage}
\usepackage{notoccite}
\usepackage{amsfonts} 
\usepackage{setspace}
\usepackage{amssymb}
\usepackage{fancyhdr}
\usepackage{authblk}

\gdef\@fpheader{Prepared for submission to IEEE Microwave Magazine}

\usepackage[normalem]{ulem}
\usepackage[style=ieee,backend=biber,hyperref=auto]{biblatex}
 \useunder{\uline}{\ul}{}
 
\date{}

\title{Microwave Technologies in Experiments for Detection of Dark Matter Axions}

\author[1]{Jose R. Navarro-Madrid\thanks{corresponding author: joser.navarro@upct.es}}
\author[2]{Jose María García-Barceló\thanks{ jmgarcia@mpp.mpg.de}}
\author[1]{Alejandro\:Díaz-Morcillo\thanks{ alejandro.diaz@upct.es}}
\affil[1]{Department of Information Technologies and Communications, Universidad Politécnica de Cartagena, 30202 - Cartagena, Spain}
\affil[2]{Max-Planck-Institut für Physik (Werner-Heisenberg-Institut), Boltzmannstr. 8, 85748 - Garching bei M\"{u}nchen, Germany}

\begin{document}
\maketitle

\noindent{\small\scshape\@fpheader}\par
\doublespacing

\section{Introduction}
\begin{figure}[ht]
    \centering
    \includegraphics[scale=0.5]{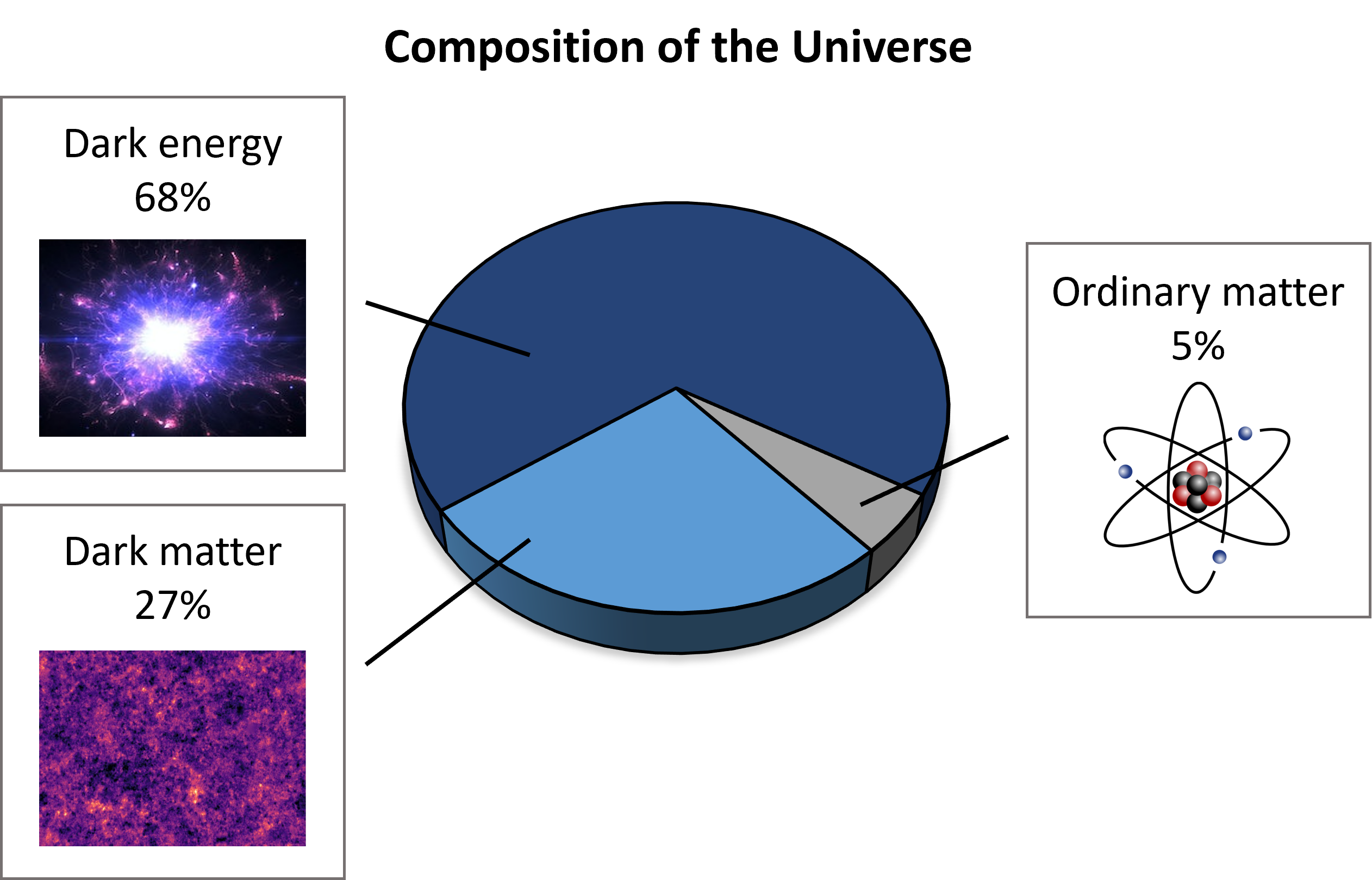}
    \caption{Universe composition graph. The major part is composed of dark energy which would be present throughout space and produces a pressure that tends to accelerate the expansion of the universe. The $27$~$\%$ corresponds to the dark matter, whose existence has been theorized by different cosmological phenomena. Finally, the rest is conformed by ordinary matter.}
    \label{fig:composition_univ}
\end{figure}

What is our universe made of? This is one of the great questions that, at present, has not been fully resolved. Thanks to advancements in technology over the past century, scientists have gained access to previously unattainable insights into the evolution of the universe and its composition, estimating that the universe would be made up of only a $5$~$\%$ of ordinary matter (the kind of matter that we can interact with). The rest is composed of two other unknown elements: dark energy ($68$~$\%$) and dark matter (DM, $27$~$\%$). A distribution of the universe composition is depicted in Figure~\ref{fig:composition_univ}. Astrophysical observations and studies carried out by F. Zwicky and others led to the hypothesis of the existence of DM \cite{Zwicky}. Its nature remains elusive, as it does not interact significantly with electromagnetic (EM) waves and therefore cannot be observed directly. Its presence is inferred through its gravitational effects on the formation and evolution of large-scale cosmic structures, such as galaxies and galaxy clusters.

First proposed to solve certain problems in quantum chromodynamics \cite{Wilczek1977,weinberg1978}, the axion also has some characteristics that make it suitable for composing DM among other candidates. Its mass would be extremely low, a property that makes it virtually undetectable through EM and strong interactions. Moreover, during the early stages of the universe, it is postulated that axions were produced in abundance due to the breaking of the Peccei-Quinn symmetry \cite{Pecceib1977,Peccei1977}. This massive production in the cosmic past would contribute to the present density of DM and explain its presence over time.

Axions are a subject of intense study, and efforts in the scientific community to test their existence experimentally have been made since 1983, when P. Sikivie proposed two experiments to detect axions from different sources \cite{Sikivie1983}. These experiments take advantage of the interaction between the axion field and the EM fields, where axions are converted into photons under a static magnetic field. This physical phenomenon is the so-called inverse Primakoff effect \cite{Primakoff1951}, and the expression that relates the frequency of the generated photons and the axion mass is
\begin{equation}
    f \approx \frac{m_a c^2}{h},
\end{equation}
being $c$ the speed of light in vacuum and $h$ the Plank constant. Currently, there are three types of axion experiments that make use of the inverse and direct Primakoff effect, being haloscope experiments the only ones that can detect DM axions. But the main obstacle to detecting the axion is that its mass cannot be predicted from theory, and it must be experimentally determined by sweeping a very broad energy range. Exhaustive reviews of axion searches can be found in \cite{Irastorza2018} and \cite{Diluzio2020}. If the axion is detected, the nature of dark matter will finally be unveiled.

---------------------------------------------------------------------

\textbf{SIDEBAR 1: Axion detection experiments}

\begin{figure}[ht]
\centering
\begin{subfigure}[b]{0.85\textwidth}
         \centering
         \includegraphics[scale=0.55]{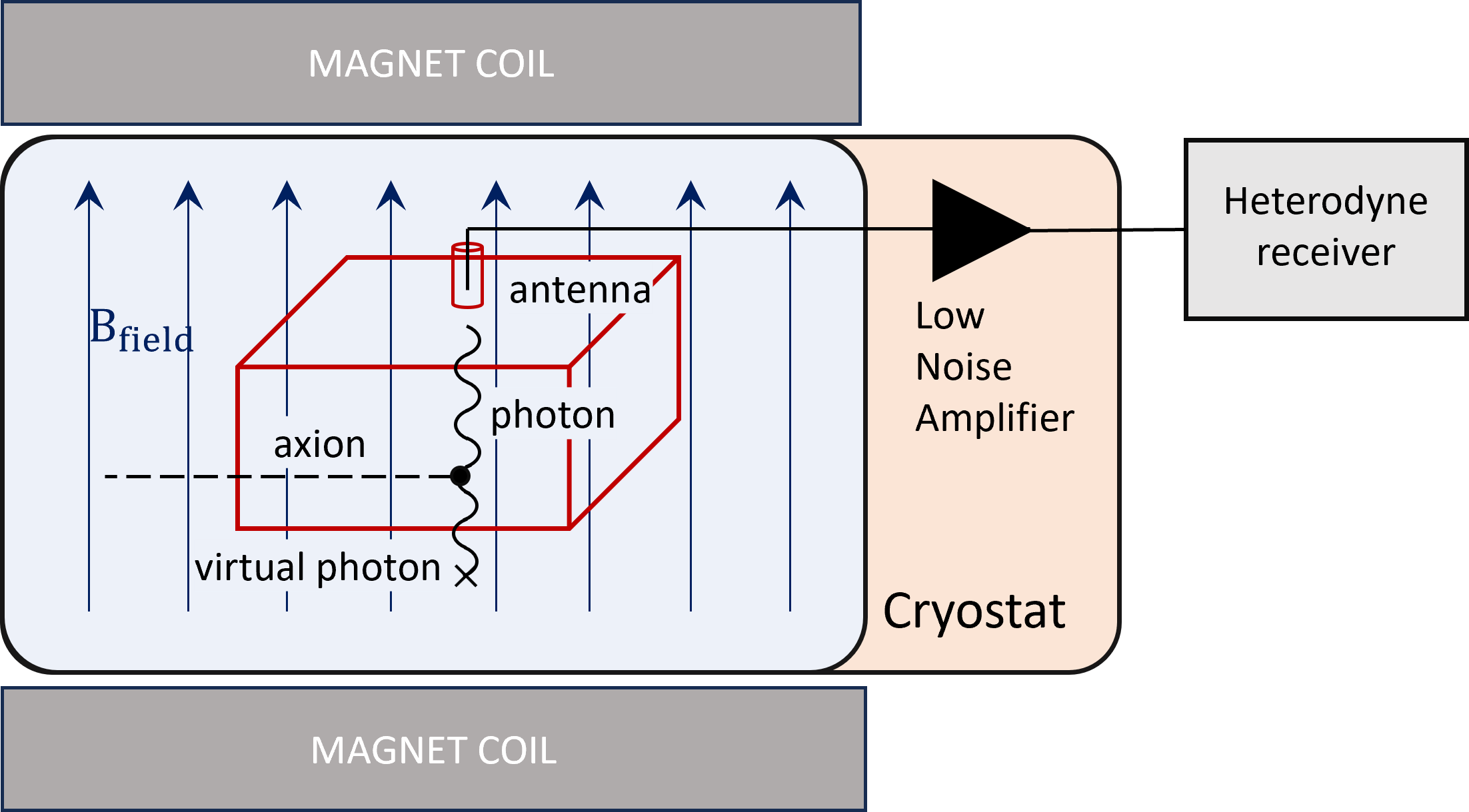}
         \caption{}
         \label{fig:haloscope_scheme}
\end{subfigure}
\hfill \medskip
\begin{subfigure}[b]{0.85\textwidth}
         \centering
         \includegraphics[scale=0.55]{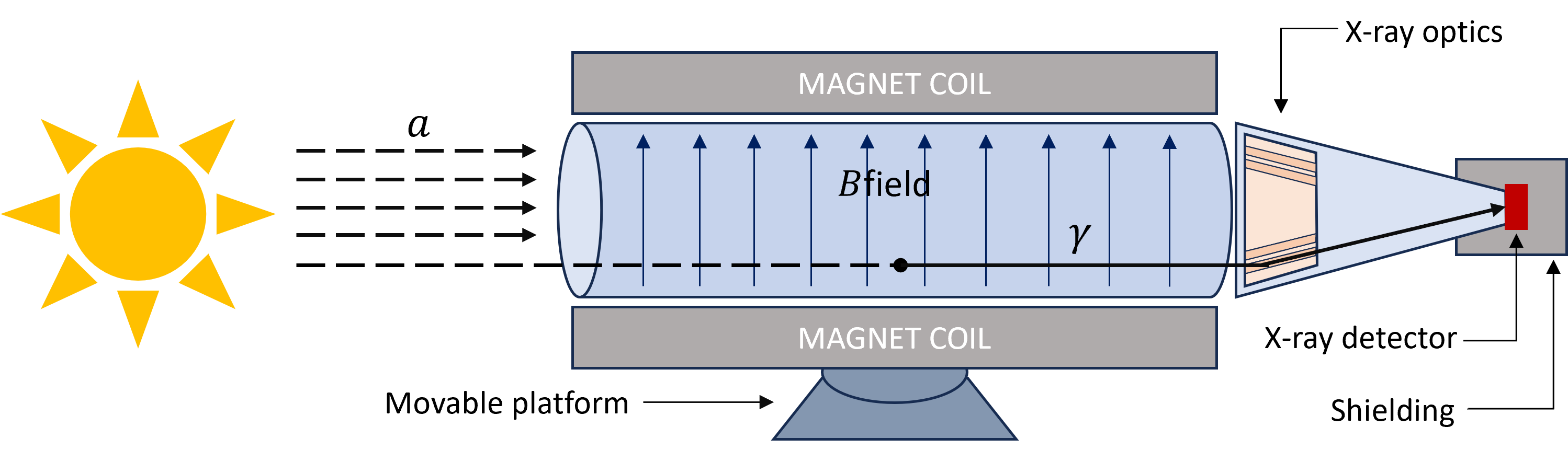}
         \caption{}
         \label{fig:helioscope_scheme}
\end{subfigure}
\hfill \medskip
\begin{subfigure}[b]{0.75\textwidth}
         \centering
         \includegraphics[scale=0.6]{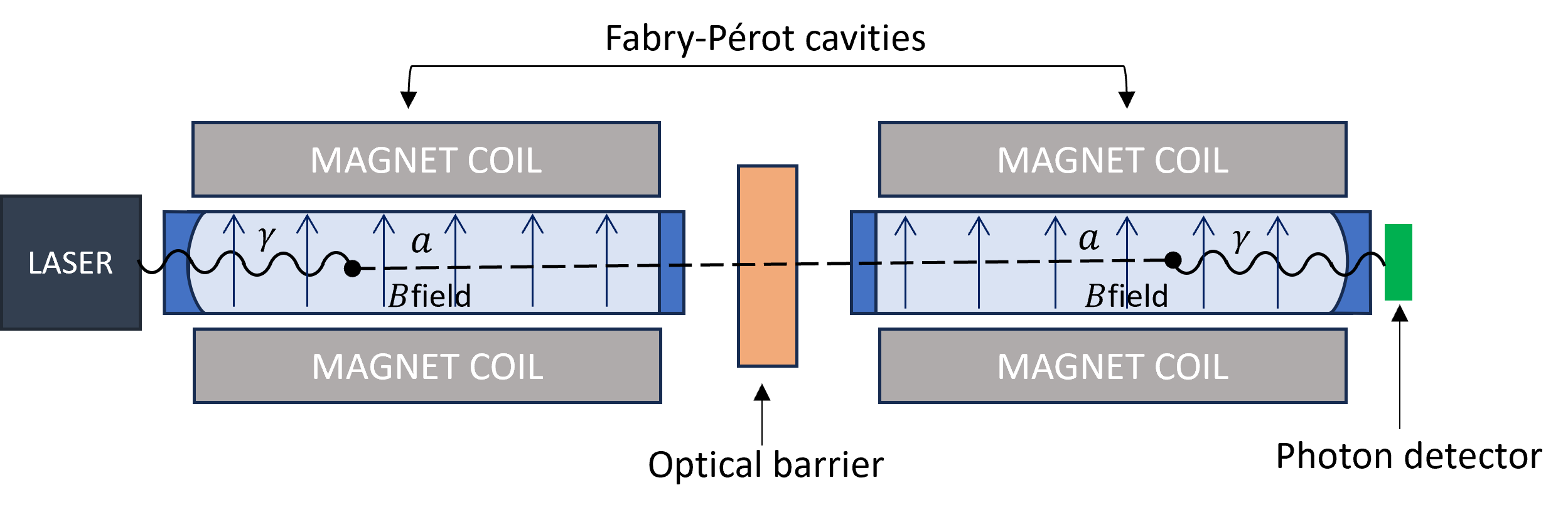}
         \caption{}
         \label{fig:lws_scheme}
\end{subfigure}
\renewcommand{\thefigure}{S\arabic{figure}}
\caption{Experiment scheme showing details of the detection principle for (a)~haloscopes, (b) helioscopes, and (c) LSWs.}
\label{fig:experiment_schemes}
\end{figure}

\textbf{Haloscopes:} An axion haloscope is a physical device whose objective is to detect and enhance EM signals induced by DM axions in our galactic halo. Typically, they are placed in high magnetic field magnets working at cryogenic temperatures (a scheme is shown in Figure~\ref{fig:experiment_schemes} (a)). Due to the dependence of the axion mass on frequency, the characteristics and dimensions of the haloscope will depend on the mass range where the search is being performed, which will involve the use of different technologies, such as microwave resonant cavities, lumped LC resonators \cite{ADMX.SLIC,ABRACADABRA}, and broadband reflectors \cite{Broadband, BREAD_THz}.

\textbf{Helioscopes:} The aim of these experiments is to detect the abundant flux of
axions emitted from our sun based on the axion-photon conversion, using a dipole
magnet directed towards the sun (see scheme in Figure \ref{fig:experiment_schemes} (b)). When the sun’s nuclear fission processes create
X-rays that scatter off electrons and protons in the presence of high electric fields,
the sun’s core can produce axions that are sent to our planet. Therefore, axion
searches in helioscopes use highly efficient photon detectors in the X-ray region. First measurements with helioscopes were carried out at Brookhaven National Laboratory
(BNL) \cite{BNL}, then at the University of Tokyo \cite{SUMICO} and lately at the CERN Axion Solar Telescope (CAST) \cite{CAST2017}. These experiments have searched different regions, and the next new generation helioscope, the International Axion Observatory (IAXO) \cite{IAXO_2019}, is expected to be a breakthrough in solar axion detection.

\textbf{Light Shining through Walls (LSW):} Axions could also be produced in the laboratory from an artificial photon source, e.g. a laser, which introduces photons into Fabry-Pérot cavities within a superconducting dipole magnet. The axions that are produced in the first cavity propagate through an optical wall to the second cavity. Then, they are transformed back into photons with their original energy by a similar magnet system behind the wall, being collected by an optical sensor. A scheme is shown in Figure \ref{fig:experiment_schemes} (c). Some LSW experiments are the Brookhaven-Fermilab-Rochester-Trieste Collaboration \cite{LSW1992}, OSQAR \cite{OSQAR}, ALPS-I \cite{ALPS-I} and ALPS-II \cite{ALPS-II}.

---------------------------------------------------------------------

\section{General concepts of the resonant haloscope experiment}

Due to space limitations, within the possible configurations, in this article we will focus on resonant cavity haloscopes that operate in the microwave range. The primary goals of an effective axion detection apparatus involve the maximization of power arising from the axion-photon coupling, augmentation of the examined axion mass spectrum (and its scanning rate), and the optimization of haloscope sensitivity. The detected power ($P_d$) depends on the inherent axion characteristics and the experimental parameters of the cavity \cite{universe}, and it can be obtained in natural units as
\begin{equation}\label{eq:Pd}
    P_d = \kappa g_{a \gamma}^2 \frac{\rho_{DM}}{m_a}B_e^2 C V Q_l.
\end{equation}

Here, $\kappa$ represents the coupling to the external receiver ($\kappa = 0.5$ denoting the critical coupling scenario), $g_{a \gamma}$ the axion-photon coupling, $\rho_{DM}$ the dark matter density, $m_a$ the mass of the axion particle, $B_e$ the magnetostatic field (provided by the employed magnet), $C$ the form factor of the haloscope, $V$ the volume of the cavity, and $Q_l$ its loaded quality factor. The form factor ($C$), quantifying the alignment between the external magnetic field and the dynamic electric field of the cavity induced by the axion-photon conversion, can be mathematically expressed as
\begin{equation} \label{eq:formfactor}
     C = \frac{|\int_V \vec{B_e} \cdot \vec{E}\:dV|^2}{\int_V |\vec{B_e} |^2 dV  \int_V \varepsilon_r |\vec{E}|^2\:dV},
\end{equation}
where $\varepsilon_r$ is the relative electrical permittivity within the cavity. An indicator of the haloscope sensitivity lies in the detectable axion-photon coupling, achievable for a specified signal-to-noise ratio $S/N$, and can be obtained as
\begin{equation} \label{eq:gag}
    g_{a \gamma} = \left( \frac{\frac{S}{N} k_B T_{sys}}{\kappa \rho_{DM} C V Q_l}\right)^{\frac{1}{2}}\frac{1}{B_e}\left(\frac{m_a^3}{Q_a \Delta t}\right)^{\frac{1}{4}},
\end{equation}
where $k_B$ is the Boltzmann constant, $T_{sys}$ the system noise temperature, $\Delta t$ the time window during which data is acquired, and $Q_a$($\approx10^6$) the axion quality factor, as detailed in \cite{universe}.

The tuning of the resonant frequency within a haloscope stands as a critical attribute, given the lack of knowledge of the axion mass value. The data acquisition strategy involves systematically scanning through a defined mass range across the spectrum, necessitating a procedure for frequency shifting. The scanning rate $\frac{dm_a}{dt}$, commonly employed as a metric to gauge haloscope performance, is derivable from eq.~\ref{eq:gag} as outlined in \cite{universe}, obtaining
\begin{equation}\label{eq:scanning_rate}
     \frac{dm_a}{dt} = Q_a Q_l \kappa^2 g_{a \gamma}^4 \left( \frac{\rho_{DM}}{m_a} \right)^2 B_e^4 C^2 V^2 \left(\frac{S}{N} k_B T_{sys} \right)^{-2}.
\end{equation}

\begin{figure}[ht]
    \centering
    \includegraphics[scale=0.75]{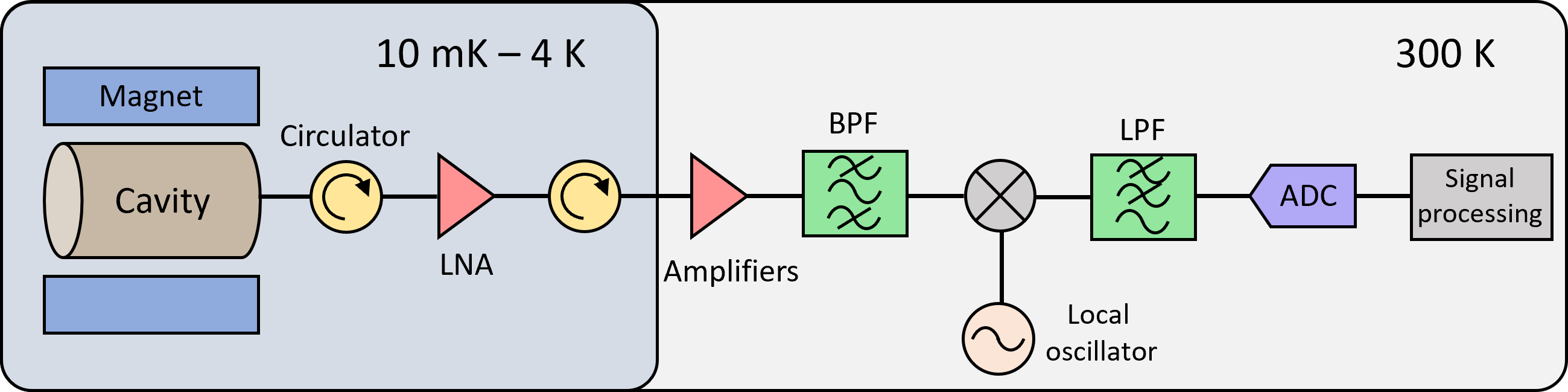}
    \caption{Schematic of a typical readout chain in haloscope experiments. Low noise components are used in the cryogenic stage, while other devices like filters and mixers are employed at room temperature.}
    \label{fig:readout_chain}
\end{figure}

Finally, the detection setup of the experiment is composed of common devices in microwave engineering. A schematic of the readout chain is shown in Figure \ref{fig:readout_chain}.

---------------------------------------------------------------------

\textbf{SIDEBAR 2: Cavity configuration depending on the magnet}

\begin{figure}[h!]
\centering
\begin{subfigure}[b]{0.3\textwidth}
         \centering
         \includegraphics[scale=0.8]{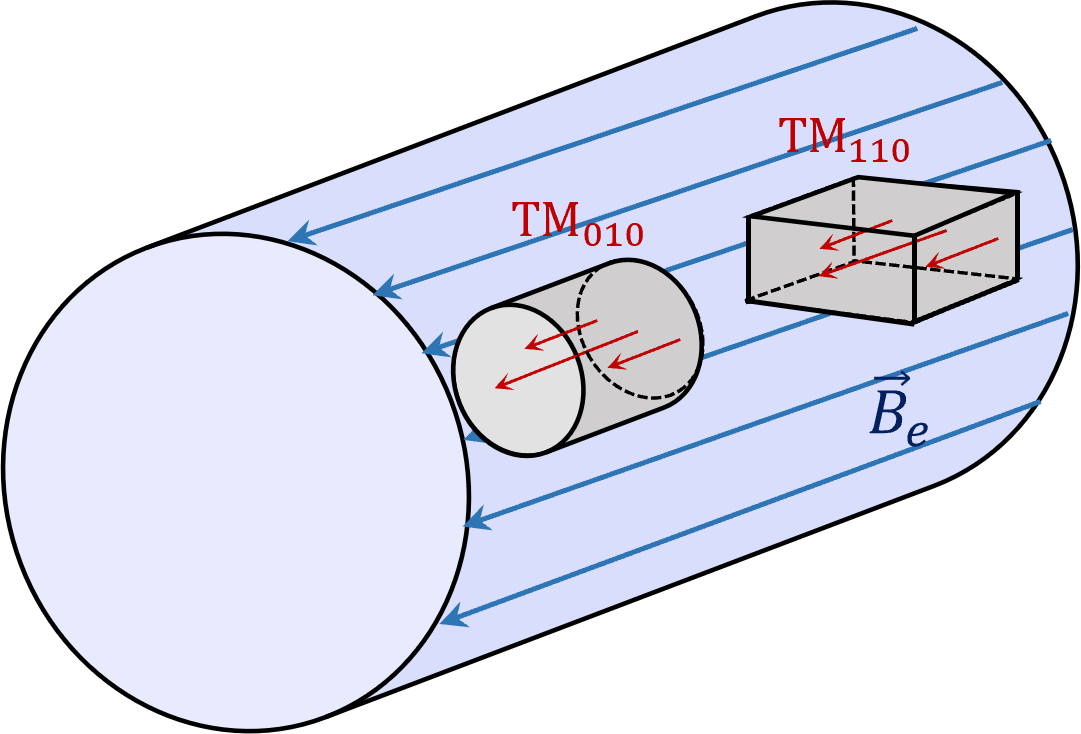}
         \caption{}
         \label{fig:RecAndCylCavities_Solenoid}
\end{subfigure}
\hfill
\begin{subfigure}[b]{0.6\textwidth}
         \centering
         \includegraphics[scale=0.8]{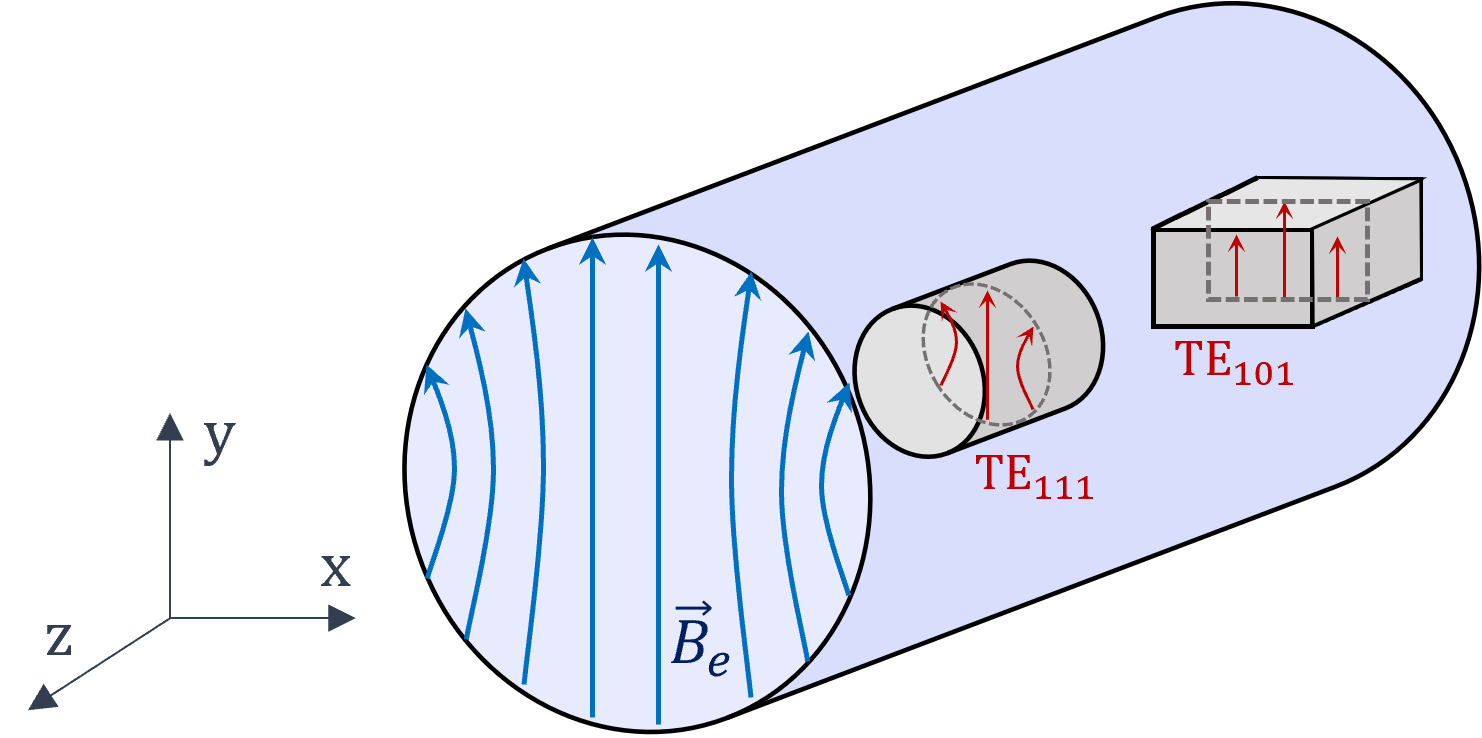}
         \caption{}
         \label{fig:RecAndCylCavities_Dipole}
\end{subfigure}
\renewcommand{\thefigure}{S\arabic{figure}}
\caption{Orientation of rectangular and cylindrical cavities in (a) solenoid and (b) dipole magnets. External cylinders represent the magnet bores. Blue lines show the external magnetic field distribution in the magnets, while red lines depict the electric field of the chosen resonant modes in each cavity.}
\label{fig:RecAndCylCavities_Solenoid_Dipole}
\end{figure}

The orientation of the magnetic field will determine the resonant mode to be used for axion data acquisition since its electric field should be as parallel as possible to the magnetostatic field for increasing the form factor ($C$). Numerous haloscope experiments, such as ADMX \cite{ADMX_status}, HAYSTAC \cite{HAYSTAC} or CAPP \cite{CAPP}, traditionally employ ad hoc solenoid magnets that generate an axial magnetic field. Consequently, the cavity configuration involves a cylindrical cavity aligned parallel to the magnet bore axis, optimizing $C$ by facilitating the excitation of a $\textrm{TM}_{010}$ mode characterised by an axial electric field. This is the preferred type of magnet, and most of the current experiments use it. A schematic description of the magnetic field arrangement and the modes used in this type of magnet can be seen in Figure~\ref{fig:RecAndCylCavities_Solenoid_Dipole} (a). Other structures with tall rectangular geometries could be used in solenoid magnets. In this case, the chosen resonant mode is the $\textrm{TM}_{110}$, which produces a vertical-polarized electric field \cite{VolumeRec}.

In contrast with solenoid magnets, the less-used dipole magnet \cite{CAST2017,BabyIAXO2021} produces a transverse magnetic field. For compatibility with such magnets, a more suitable cavity shape is the rectangular configuration, where the $\textrm{TE}_{101}$ mode features a vertically polarized electric field. However, their geometrical counterpart, cylindrical cavities, can also be used for this type of magnet, being the $\textrm{TE}_{111}$ mode the one whose electric field is parallel to the dipole magnetic field \cite{VolumeCyl} (see Figure~\ref{fig:RecAndCylCavities_Solenoid_Dipole} (b)).

---------------------------------------------------------------------

\section{Experimental objectives and associated technologies}

Taking into account the extremely low interaction between axions and photons, very small detection power is expected, in the order of $10^{-22} \sim 10^{-24}$ W, depending on the working frequency. This points out the necessity of optimizing the operational parameters involved in the experiment. From eq.~\ref{eq:scanning_rate} we can identify those parameters imposed by nature ($g_{a \gamma}$, $m_a$, $\rho_{DM}$), those that are given by the magnet ($B_e$) and partially by the cryostat ($T_{sys}$), and those that can be optimized through different microwave technologies (cavity volume, quality and form factors, coupling between transmission line and cavity, and noise added by the read-out chain). This gives us the following Figure of Merit to quantify the resonator performance:
\begin{equation} \label{eq:FoM}
    \textrm{FoM} = Q_0 V^2 C^2.
\end{equation}

Since $C$ depends mainly on the operational cavity mode and the external magnetic field, the parameters to be optimized in the cavity design are $Q_0$ and $V$. In this section, we identify some techniques currently used or proposed for improving these parameters.

\subsection{Volume}

When the magnet employed in the experiment is much bigger than the resonant cavity, typically for high frequencies, it is possible to increase the volume of the cavity in order to improve the sensitivity and, ultimately, the resonator FoM. But if not made properly, the increase in the cavity size will decrease its resonant frequency, moving away from our frequency target. Different strategies have been proposed for increasing the volume without impacting the resonant frequency.

Increasing the size of the cavity in that direction which does not affect the resonant frequency is the simplest way to gain volume. Thus, when working with solenoids and rectangular or cylindrical cavities, the $\textrm{TM}_{110}$ or $\textrm{TM}_{010}$ mode, respectively, allows to increase the length of the cavity ($z$-direction in Figure~\ref{fig:RecAndCylCavities_Solenoid_Dipole}). Moreover, when one of the dimension with EM field variation is much bigger than the other one, the resonant frequency is set by the smallest dimension, and the large one can be increased with negligible effect on the resonant frequency. This is the case, for instance, of rectangular geometries with mode $\textrm{TE}_{101}$ in dipole magnets, when the length ($d$) is much larger than the width ($a$), leading to
\begin{equation}
    f_{TE_{101}} = \frac{c}{2} \sqrt{\frac{1}{a^2}+\frac{1}{d^2}} \approx \frac{c}{2a}.
\end{equation}

The limit of this increase is given by the effect of mode clustering when the cavity becomes very large electrically. This clustering hinders the identification of the resonance and even can decrease the form factor around the resonance peak since the energy from the photons is divided between the axion mode and the closest neighbour mode, provided that that mode is excited by the photon. A complete analysis of this limitation is presented in \cite{VolumeRec} for rectangular cavities and \cite{VolumeCyl} for cylindrical cavities.

An alternative or even complementary idea for increasing the volume is the use of multiple identical cavities stacked within the magnet, whose signal is coherently summed in a power combiner. In that way, the total detected power is the sum of the power of each cavity, and, so, the effective volume is the sum of individual volumes. This strategy has been proposed in \cite{CAPP_multiple} for $6$~GHz and \cite{CADEX} for W-band frequencies. Fortunately, it is assumed from theoretical models that the axion de Broglie wavelength, and so that of the photon produced from it, is $\sim 100$~meters. Therefore, for experiments with a small or medium size, we can consider that the photons produced in the different cavities have the same phase. The main drawback here is that care must be taken with the path from each cavity to the power combiner in order to guarantee the same length and, therefore, the same phase shift in the combiner.

\begin{figure}[ht]
\centering
\begin{subfigure}[b]{0.45\textwidth}
         \centering
         \includegraphics[width=1\textwidth]{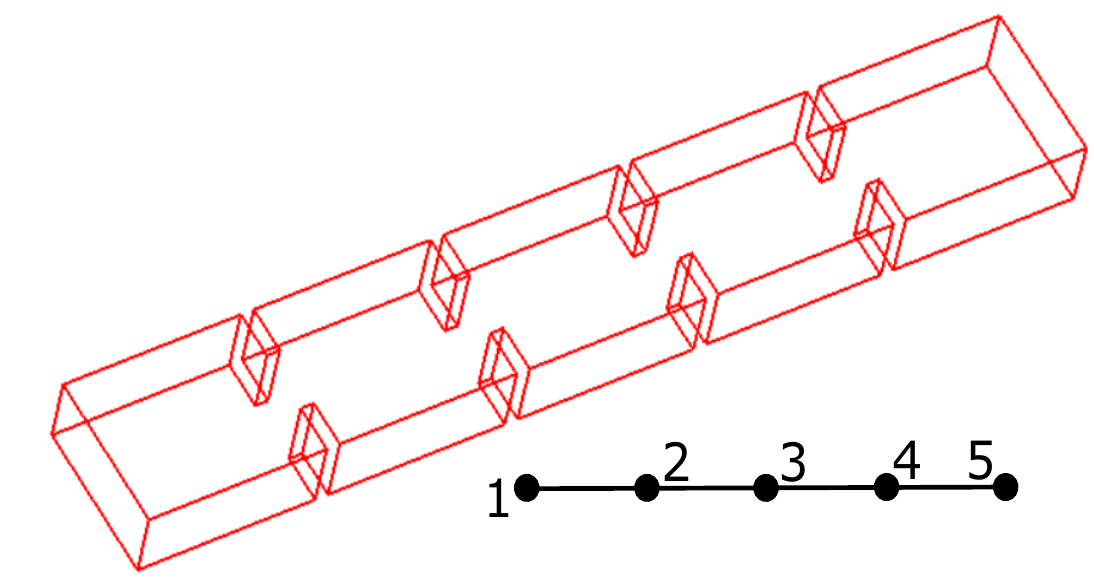}
         \caption{}
         \label{fig:1D_5cav}
\end{subfigure}
\hfill
\begin{subfigure}[b]{0.45\textwidth}
         \centering
         \includegraphics[width=1\textwidth]{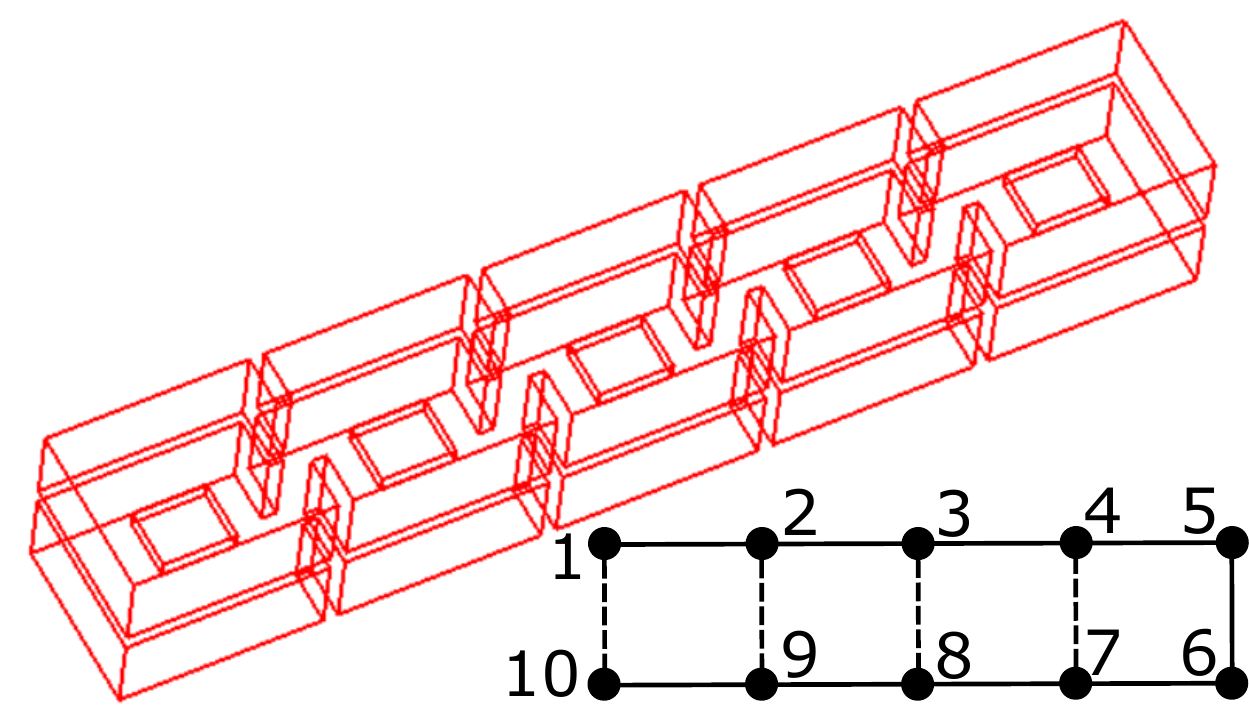}
         \caption{}
         \label{fig:2Dvertical}
\end{subfigure}
\hfill
\begin{subfigure}[b]{0.5\textwidth}
         \centering
         \includegraphics[width=1\textwidth]{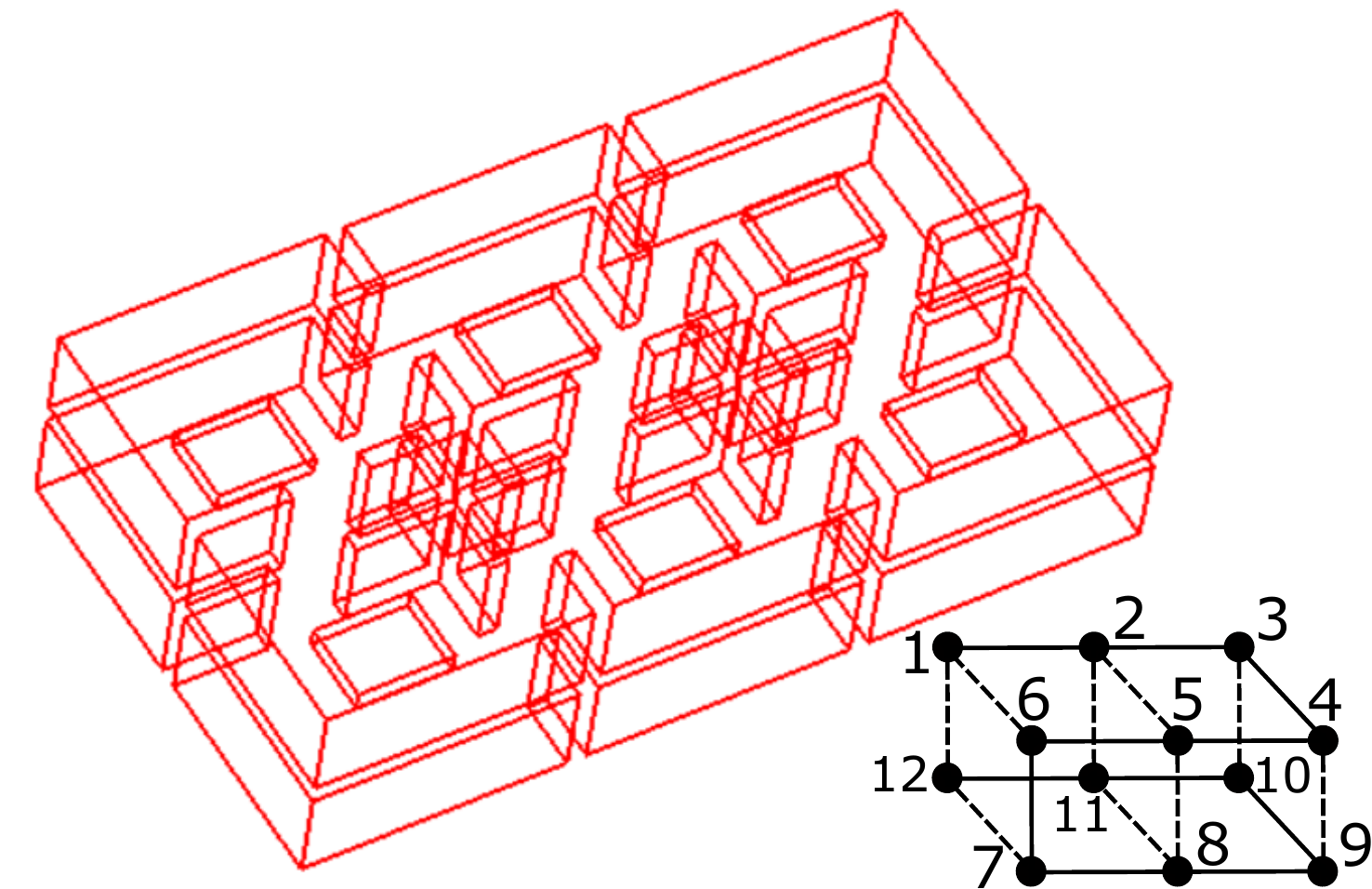}
         \caption{}
         \label{fig:3D2vertical2horizontal}
\end{subfigure}
\hfill
\begin{subfigure}[b]{0.35\textwidth}
         \centering
         \includegraphics[width=1\textwidth]{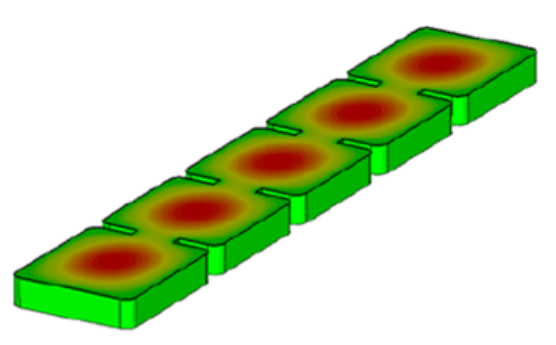}
         \caption{}
         \label{fig:5cav_Cmax}
\end{subfigure}
\caption{Examples of rectangular multicavity haloscopes of (a) 1D, (b) 2D, and (c) 3D type. (b) and (c) are taken and modified from \cite{VolumeRec}. (d) Electric field pattern (vertical polarization) for the configuration mode $\textrm{TE}^{+++++}_{101}$ which produces the highest form factor in the multicavity of (a). The red fields denote positive
levels, and the green fields zero. Taken and modified from \cite{VolumeRec}.}
\label{fig:1D_2D_3D_multicavity}
\end{figure}

In order to avoid this problem, it is possible to obtain a coherent addition of signals using only one resonant device by means of the multicavity concept. This is a well-known idea from cavity filter design \cite{Cameron} and it is adapted for DM axion detection in \cite{RADES2018,universe,CAPPmulticav}. Depending on the number of axes used to stack the subcavities, the multicavity will be of type 1D, 2D, or 3D, as Figure~\ref{fig:1D_2D_3D_multicavity} depicts. The increase in volume can be enhanced when complementing multicavity and large cavities, that is, a multicavity made up of large cavities. Again, the limitation in volume comes from the clustering of cavity modes and multicavity configurations. A complete analysis of these limits can be found in \cite{VolumeRec} and \cite{VolumeCyl} for rectangular and cylindrical multicavities, respectively. There it is shown that it is possible to enhance the FoM with a proper multicavity design up to $\sim175$~times regarding a standard cavity.

\begin{figure}[ht]
    \centering
    \includegraphics[scale=0.7]{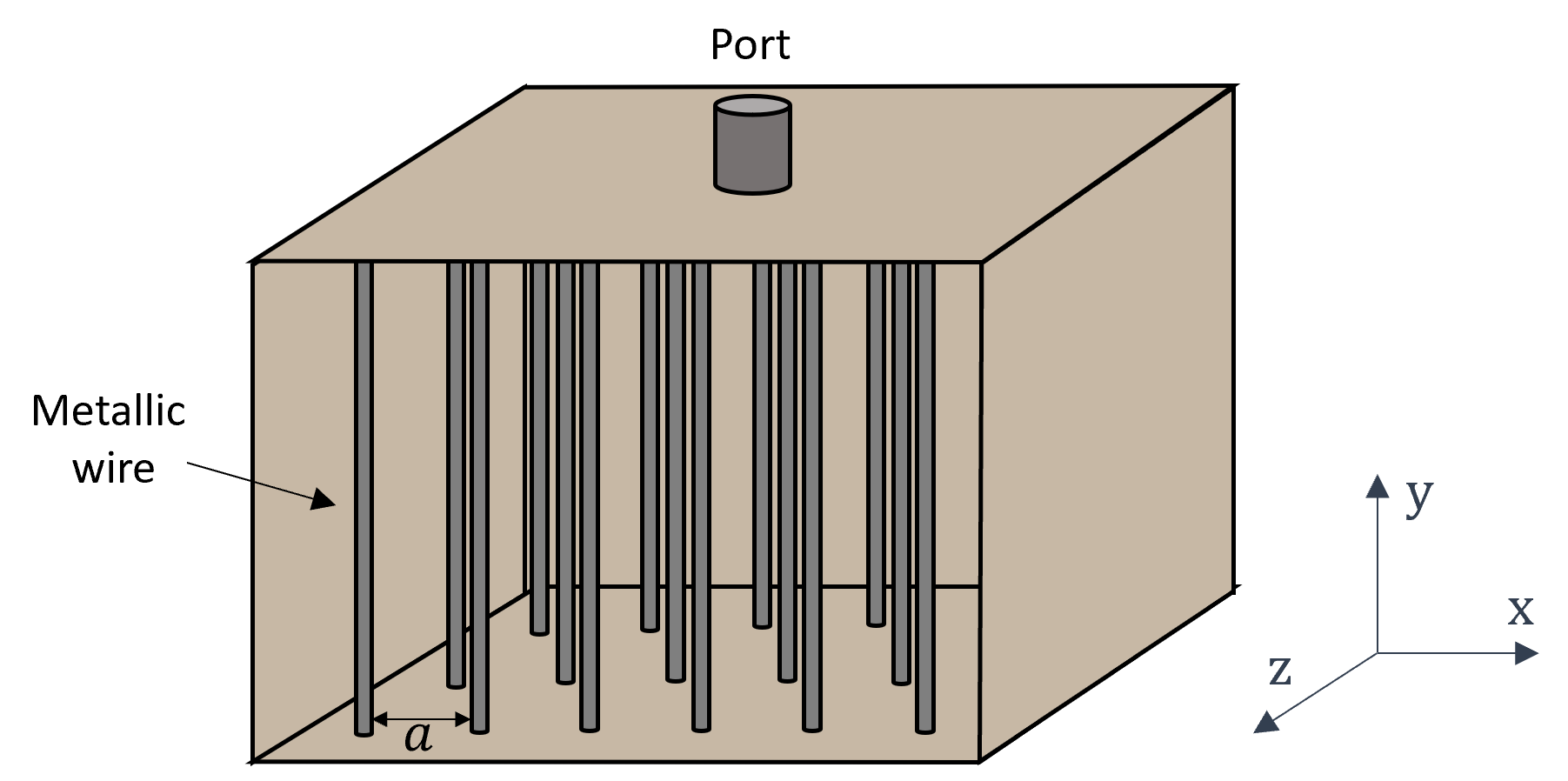}
    \caption{Wire medium cavity scheme. The distance between wires ($a$) is directly related to the resonant frequency of the operation mode, allowing the volume to be increased while maintaining a high resonant frequency.}
    \label{fig:wire_array}
\end{figure}

Another interesting concept proposed for increasing the volume is the metamaterial or plasma haloscope, based on the axion-plasmon interaction, which occurs when a plasma medium is present \cite{Plasma}. To create this medium, haloscopes have been developed where the inner volume is filled with metallic wires that create a medium with plasma characteristics, where the resonant frequency depends on the distance between these wires independently of the physical dimensions of the cavity (as is the case with conventional haloscopes). This allows the search for axions at higher frequencies while maintaining a large volume \cite{Wire_array}. Figure \ref{fig:wire_array} depicts this wire array cavity concept.

Although multicavity and metamaterial concepts are very useful for increasing the volume, their main drawback is the complexity of the structure, not only from the manufacturing point of view, especially when working at high frequencies ($>30$~GHz), but mainly for the difficulty of implementing proper tuning mechanisms within them.

\subsection{Quality factor}

The straightforward way to increase the cavity quality factor is by increasing the electrical conductivity of the inner walls or any other metallic part inside the cavity. Usually, haloscope cavities are made of high-purity copper or copper-plated non-magnetic stainless steel. It is important to use non-magnetic metals in order to avoid damage to the cavity in the case of quench events in the magnet. An estimated conductivity of $2\times 10^9$~S/m for copper at cryogenic temperature allows $Q_0$ in the order of $10^5$ for a hundred of MHz or $10^4$ for a few or dozens of GHz.

In order to increase these values, superconductor materials must be used. Type I superconductors have produced very high $Q_0$ ($\sim 10^{11}$) in cavities for particle accelerators \cite{Superconducting}, but they cannot be used in the haloscope experiment because of the strong magnetic field that permeates the cavity. Fortunately, type II superconductors, based on rare earth oxides, are magnetic-resilient and are able to achieve high, although more moderate, $Q_0$ in the order of $10^6$, as reported in \cite{Superconducting2}. It is important to note that in this case, slight losses due to gaps between the superconductor films or seams/slits in the cavity junctions can spoil the quality factor completely. Therefore, both new coating techniques and new cavity geometry and manufacturing processes are of key importance to overcome this problem.

A completely different way to improve the quality factor is employed in \cite{Quax_Q} by the use of low-loss dielectric inside the cavity and choosing a higher mode as axion mode. Although a higher-order mode and the use of dielectric reduce significantly the form factor (eq.~\ref{eq:formfactor}), the resonant frequency increases, and if there is enough room in the magnet bore, the volume of the cavity can be increased to get the target frequency. In this way, the factor $C^2V^2$ does not change too much, and we get a higher $Q_0$ due to the fact that the currents in the walls are smaller due to the dielectric confinement and the high-order mode.

---------------------------------------------------------------------

\textbf{SIDEBAR 3: Tuning systems} 

A complete haloscope experiment requires the exploration of a target frequency range corresponding to an axion mass range. In a typical resonant haloscope experiment, the objective is to be sensitive to the axions in a frequency range (typically $1-10$~$\%$) around a central frequency. This frequency sweep is achieved by means of a tuning mechanism that is able to change the resonant frequency of the cavity, minimizing the impact on other important parameters such as the form factor or the quality factor.

\begin{figure}[ht]
\begin{minipage}{.49\linewidth}
\centering
\subfloat[]{\label{main:a}\includegraphics[scale=.65]{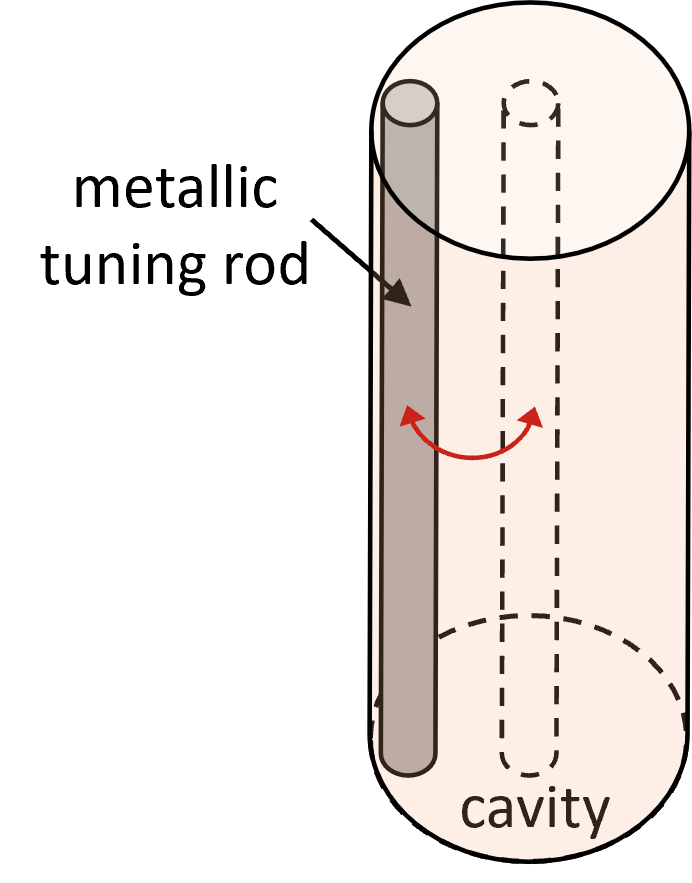}}
\end{minipage}
\begin{minipage}{.49\linewidth}
\centering
\subfloat[]{\label{main:b}\includegraphics[scale=.65]{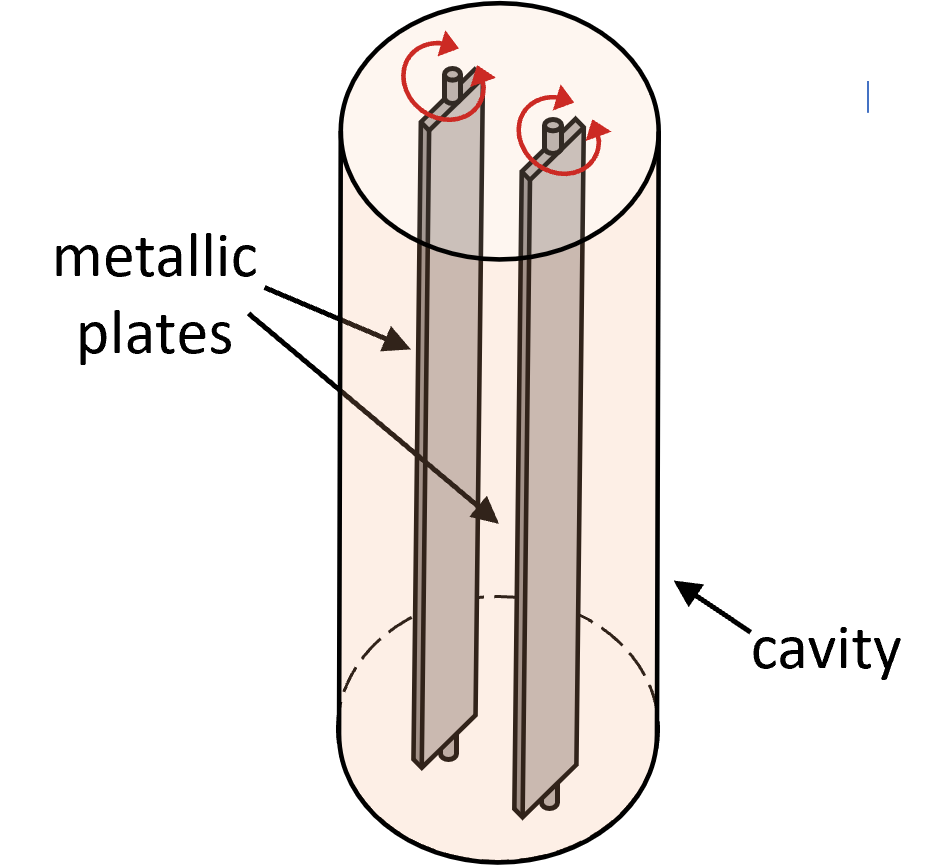}}
\end{minipage}\par\medskip
\begin{minipage}{.49\linewidth}
\centering
\subfloat[]{\label{main:c}\includegraphics[scale=.7]{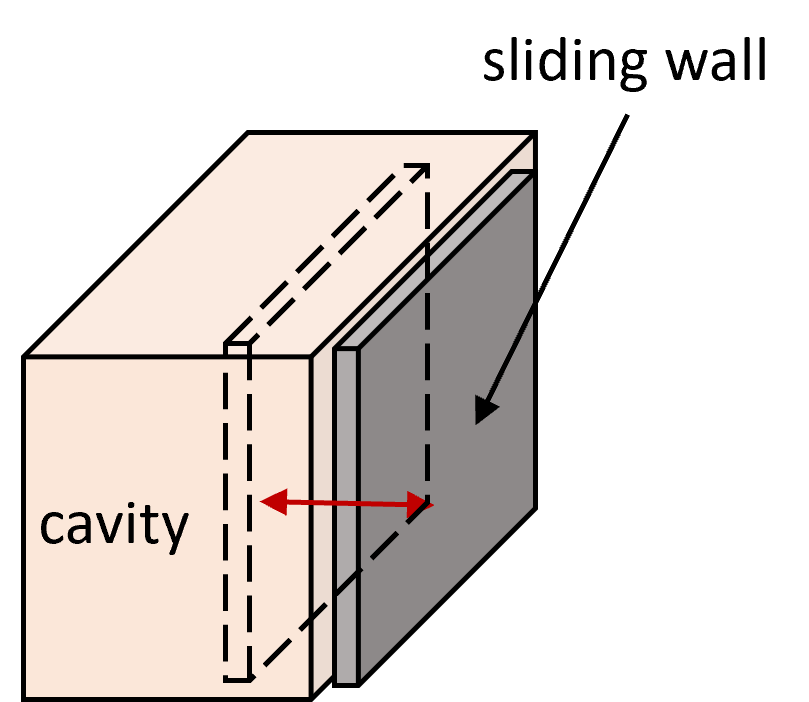}}
\end{minipage}
\begin{minipage}{.49\linewidth}
\centering
\subfloat[]{\label{main:d}\includegraphics[scale=.7]{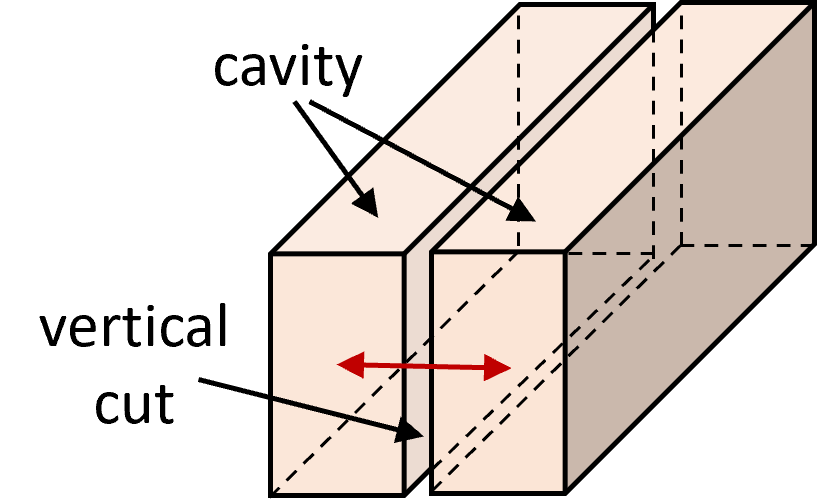}}
\end{minipage}
\renewcommand{\thefigure}{S\arabic{figure}}
\caption{Different tuning mechanical procedures in cavity haloscopes: (a) metallic rod, (b) rotating plates, (c) sliding wall, and (d) vertical cut. In (a) an off-centered angular movement is applied, in (b) a centered rotational motion, and in (c) and (d) a longitudinal movement.}
\label{fig:mechanical_tuning}
\end{figure}

\begin{figure}[ht]
    \centering
    \includegraphics[scale=0.7]{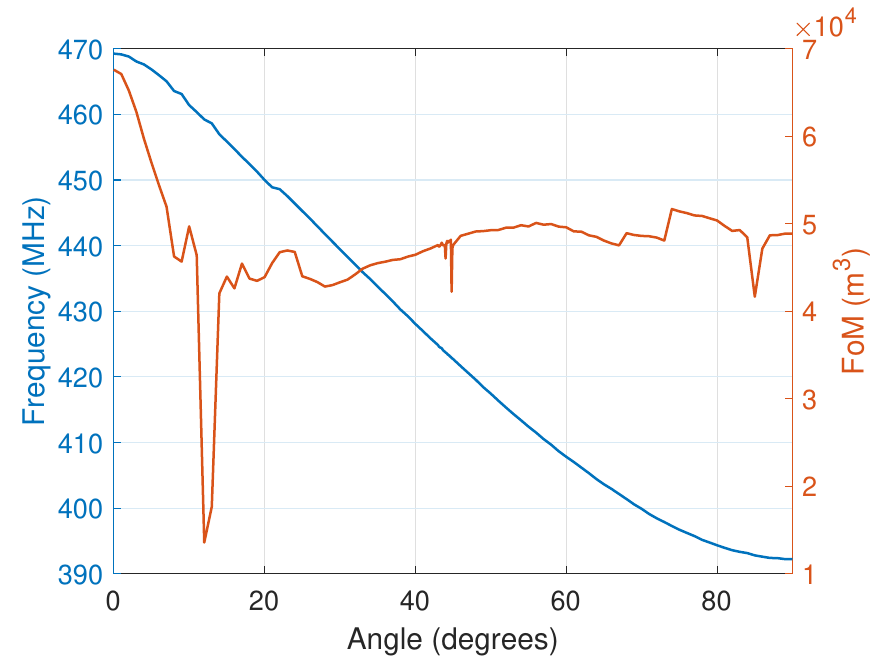}
    \caption{Resonant frequency (red line) and FoM (blue line) versus rotating plates angle from one of the cavities described in \cite{BABYIAXO_RADES}.}
    \label{fig:UHF_tuning}
\end{figure}

Mechanical tuning is widely used in haloscopes because of the greater ease of changing the electric field distribution compared to electrical tuning. However, it requires the use of auxiliary systems, such as motors or piezoelectrics, that occupy space inside the magnet bore. In addition, these devices must be compatible with working at cryogenic temperatures and under strong magnetic fields. So far, the most commonly used mechanism is to move a metallic cylinder inside the cavity \cite{ADMX_tuning,HAYSTAC} from the edge to the centre, changing the EM field distribution of the $TM_{010}$ and thus its resonant frequency (see Figure~\ref{fig:mechanical_tuning} (a)). For different resonant modes, other techniques have been proposed, such as the use of rotating plates inside the cavity for $TE_{111}$ in \cite{BABYIAXO_RADES} (see Figure~\ref{fig:mechanical_tuning} (b)). For rectangular cavities, the displacement of one of the cavity walls allows to change one of its dimensions depending on the EM mode of operation, affecting the resonant frequency (see Figure~\ref{fig:mechanical_tuning} (c)) \cite{CADEX,ORGAN_slidingwall}. These moving parts must incorporate mechanisms to avoid radiation, such as the use of gaskets to improve metal contact or radio-frequency chokes. Moreover, there is the possibility to perform tuning based on 'vertical cut', i.e. separating two halves of a cavity and thus varying its width (see Figure~\ref{fig:mechanical_tuning} (d)). A study of this type of tuning can be found in \cite{vertical_cut}. In Figure~\ref{fig:UHF_tuning}, the obtained tuning range for a cavity based on rotating plates and the FoM are shown. Minima in the FoM are due to mode crossings, and those frequencies are considered as blind regions for axion detection in the experiment.

\begin{figure}[ht]
    \centering
    \includegraphics{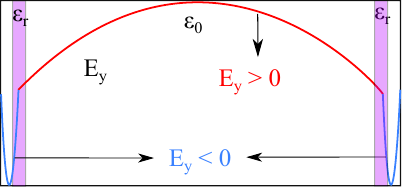}
    \renewcommand{\thefigure}{S\arabic{figure}}
    \caption{Transverse view of $E_y$ (vertical component of the E-field) inside the resonant cavity (white) with ferroelectric elements (purple). The red line denotes a positive value of the electric field, while the blue lines show a negative value. The variation in the permittivity ($\varepsilon_r$) shifts the 0-level electric field position, and hence it provides a frequency change. Taken from \cite{Ferroelectric}.}
    \label{fig:ferroelectric_tuning}
\end{figure}

In contrast to mechanical tuning, electrical tuning systems can provide a design that is more resistant to mechanical failures (e.g., by omitting movable components, which may pose a risk at cryogenic temperatures), which reinforces and complements current methods. Moreover, scaling mechanical systems at higher frequencies is challenging, due to the small dimensions of the devices. Electrical tuning for haloscopes by using ferroelectric materials has been studied in \cite{Ferroelectric}. These materials allow their permittivity to be changed by applying an external DC voltage, thus, changing the mode frequency. Figure~\ref{fig:ferroelectric_tuning} shows an scheme of a cavity with ferroelectric films for tuning. Along the same lines, another current research is tuning with a change in permeability by means of ferromagnetic materials, allowing the frequency to be controlled by the magnetic field of the magnet \cite{PhD_JMGB}.

Regardless of the tuning system, it will slightly modify the EM field distribution in the cavity and, therefore, the coupling between the transmission line and the cavity. Since the haloscope performance (eq. \ref{eq:scanning_rate}) depends on this coupling, it is mandatory to have a movable probe in order to keep the target coupling from one frequency point to the next. Paradoxically, this target is not $\kappa=\frac{1}{2}$ (critical coupling), but $\kappa=\frac{2}{3}$ (overcoupled) or even higher. A complete analysis of optimal coupling can be found in \cite{CAPP_revisiting}. For electrical coupling, a moving monopole antenna is normally used, while a rotating loop is employed for magnetic coupling \cite{BABYIAXO_RADES}. For millimeter waves, where coaxial lines are avoided due to their high losses, an iris on the waveguide connected to the cavity and a moving pin are proposed in \cite{CADEX}.

---------------------------------------------------------------------

\subsection{Noise}

Reducing the system noise is a key factor in improving the haloscope performance, so working at cryogenic temperatures is mandatory, thus reducing the thermal noise of the cavity and the first amplifier in the readout chain.

Several technologies can be used to amplify the low-expected signal. The most commonly used devices are Low Noise Amplifiers, which introduce a small noise to the signal, offering a typical system temperature of around $5-10$~K \cite{universe}. It is also possible to approach the Standard Quantum Limit (SQL) \cite{SQL} by using superconducting amplifiers, such as Superconducting Quantum Interference Devices (SQUIDs). These detectors have been applied in \cite{ADMX_SQUID}, being able to reduce the system noise temperature to $\sim 300$~mK. Other proposals that make use of Josephson Parametric Amplifiers (JPAs) \cite{Aumentado2020} or Josephson Traveling Wave Parametric Amplifiers (JTWPAs) \cite{JTWPA_1985} can be seen in \cite{JPA} and \cite{JTWPA}, respectively.

\begin{figure}[ht]
    \centering
    \includegraphics[scale=0.7]{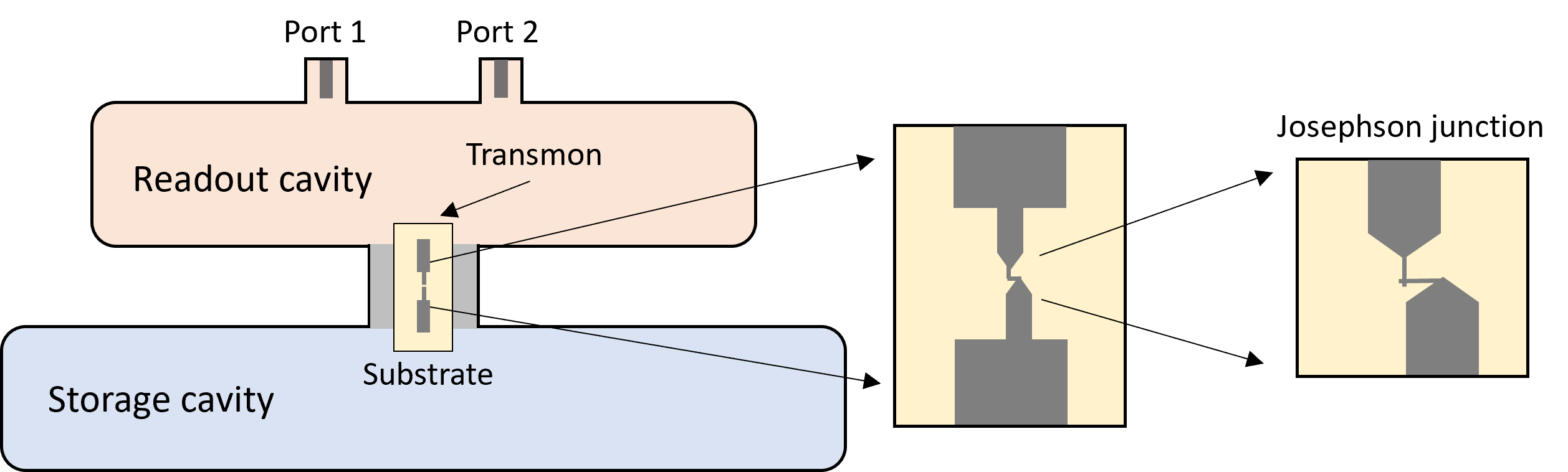}
    \caption{Scheme of a qubit-cavity system for DM detection. The dark photon-photon interaction takes place in the storage cavity and produces a change in the state of the qubit. The readout cavity sends signals to the qubit to check the current state. In the image, a zoom of the Josephson junction section is shown.}
    \label{fig:scheme_qubit}
\end{figure}

Recent proposals significantly reduce the noise by using photon counting devices, which detect individual photons and can be useful at lower temperatures ($T\:\sim\:$~mK) and high cavity $Q_0$ values, allowing the noise to go beyond the SQL. A complete performance analysis of single-photon amplifiers compared to linear amplifiers can be found at \cite{AnalysisAmplifiers}. A photon counter based on a Superconducting Transmon Qubit \cite{Transmon} has been proposed as a quantum detector for DM detection in \cite{Qubit}, in this case for dark photons, not axions, where the external magnetic field is not needed. Figure~\ref{fig:scheme_qubit} depicts the qubit-cavity system, where the qubit interacts with the storage cavity and change its state if photons are detected. For higher frequencies, the use of Kinetic Inductance Detectors (KIDs) \cite{KIDs2003} is described in \cite{CADEX}. These devices are high-quality factor superconducting resonators that absorb photons, producing a variation effect in the kinetic inductance of the resonator and enabling the detection of extremely low-power signal. The major obstacle in using quantum devices in DM axion detection is their incompatibility with large magnetic fields. Therefore, the development of new magnet-resilient quantum devices is a current research line.

\section{Conclusions}

In the last forty years, an accelerating race in the quest for the DM axion has taken place within the particle physics community. As seen, the different detection schemes make intensive use of microwave technologies, and the success of the future detection of this particle depends on the improvement of well-known techniques and the development of new ones. Microwave engineering has the opportunity to contribute to solving the dark matter mystery. From the review presented here, we can summarize a non-exhaustive list of research lines where key contributions can come from:

- Multicavity filters and metamaterial cavities for high-volume and high-Q devices.

- New electromagnetic mechanisms for resonator tuning, including ferroelectric and ferromagnetic materials.

- New magnetic field-resilient superconductor materials and new cavity geometries for them to reach very high quality factors.

- Single photon counters based on quantum devices in order to drastically reduce the noise of the readout chain. Some examples are 3D transmon qubits.

Advances in these lines will increase the probability of detecting not only the "invisible" axion but other particles such as the dark photon (another candidate to make up the dark matter) or the graviton (gravitational waves), since recently haloscope schemes have been proposed for both particle detection \cite{DarkPhotons,GW,GW2}.

\section*{Acknowledgments}
This work was performed within the RADES group. We thank our colleagues for their support. It is part of the R$\&$D projects PID2019-108122GB-C33 and PID2022-137268NBC53, funded by MICIU/AEI/10.13039/501100011033/ and by “ERDF/EU”. JMGB thanks the European Research Council grant ERC-2018-StG-802836 (AxScale project), and the Lise Meitner program “In search of a new, light physics” of the Max Planck society. 

\appendix

\printbibliography[title={Bibliography},heading=bibintoc] 

\end{document}